\documentclass[twocolumn,prl]{revtex4}
\usepackage{graphicx}    
\usepackage{dcolumn}     
\usepackage{bm}          
\usepackage{amsmath}
\begin{document}
\preprint{Circuit}

\title{Anomalous charge transport
in triplet superconductor junctions
}
\author{
Y. Tanaka$^{1,2}$ and S. Kashiwaya$^{3}$
}%
%
\affiliation{
$^1$Department of Applied Physics,
Nagoya University, Nagoya, 464-8603, Japan \\
$^2$
CREST Japan Science and Technology Cooperation (JST) 464-8603
Japan \\
%
%
%
$^3$National Institute of Advanced Industrial Science
and Technology, Tsukuba, 305-8568, Japan
}
%
\date{\today}
\begin{abstract}
Charge transport properties of a diffusive 
normal metal /triplet superconductor (DN/TS) junction
are studied based on the Keldysh-Nambu quasiclassical
Green's function formalism.
Contrary to the unconventional singlet superconductor junction case,
the mid gap Andreev resonant state (MARS) at the interface of
the TS is shown to enhance the proximity effect in the DN.
The total resistance of 
the DN/TS junction is drastically reduced and is completely
independent of the resistance of the DN in the extreme case.
Such anomalous transport accompanies a giant zero-bias peak in 
the conductance spectra and a zero-energy peak of 
the local density of states in the DN region.
These striking features manifest the presence of novel proximity effect
peculiar to triplet superconductor junctions.
\end{abstract}
\pacs{PACS numbers: 74.70.Kn, 74.50.+r, 73.20.-r}
\maketitle
Physics of superconducting junctions 
has been one of the exciting fields of solid state physics
in this decade \cite{meso}.
%
%
%
In diffusive normal metal / conventional singlet $s$-wave
superconductor (DN/CSS) junctions,
it is known that the phase coherence
between an incoming electron and an Andreev reflected hole
plays an essential role in causing
the proximity effect in the DN\cite{Beenakker,Hekking}.
It is also known that the total resistance $R$ of DN/CSS junctions
does not follow  the simple Ohm's rule,
that is, $R = R_{D}+ R_{R_{D}=0}$
where $R_{D}$ is the resistance in the DN
and $R_{R_{D}=0}$ is the resistance of the DN/CSS interface.
The reduction of $R$ becomes prominent
for low transparent junctions, where $R<R_{D}+R_{R_{D}=0}$ is satisfied
\cite{Beenakker}.
The lower limit of $R$ is given by
$R_{D}+R_{0}/2$ using the Sharvin resistance $R_{0}$,
where $R_{R_{D}=0}=R_{0}/2$ is satisfied only for perfect
transparent junctions \cite{Beenakker}.
Previous investigations of the proximity effect,
however, are limited to DN/CSS junctions.
Stimulated by the successive discovery of unconventional superconductors,
we are tempted to expect novel proximity effect
in junctions with  unconventional pairing,
$e.g.$, $p$-wave, and $d$-wave,
where  pair potentials 
have sign change on the Fermi surface.  \par
%
%
In unconventional superconductor junctions,
reflecting the internal phase of the pair potential,
charge transport becomes essentially phase sensitive.
The most dramatic effect is the appearance  of
zero bias conductance peak (ZBCP) \cite{TK95,Kashi00}
in tunneling spectroscopy due to the formation
of the mid gap Andreev resonant state (MARS)
\cite{Buch}.
The origin of the MARS is due to the anomalous interference
effect of quasiparticles at the interface, where
injected and reflected quasiparticles
feel different sign of the pair potentials \cite{Kashi00}.
It is an interesting issue to clarify the
role of the MARS on the transport
properties of  superconducting junctions.
Recently, we have developed
a theory of proximity effect, which is available for a
diffusive normal metal / unconventional singlet superconductor
(DN/USS) junction, where USS indicates anisotropic
singlet pairing like $d$-wave \cite{PRL2003,PRB2004}.
Unfortunately, however, it is revealed that 
the proximity effect and the MARS compete with each other in DN/USS junctions.
Although the interface resistance $R_{R_{D}=0}$ is reduced by the
MARS irrespective  of the magnitude of the
transparency at the interface,
the resulting $R$ is always larger than $R_{0}/2 + R_{D}$.
This is because the 
angular average of many channels at the DN/USS interface
destruct the phase coherence of the MARS and the proximity
effect (see Fig. 1).
This destructive angular average is due to
the sign change of the pair potentials felt by quasiparticles with
injection angle $\phi$ and those with $-\phi$,
where the angle $\phi$ is measured from the direction normal to the
junction interface (see Fig. 1).
However, in diffusive normal metal / triplet superconductor
(DN/TS) junctions,
we can escape from the above destructive average (see Fig. 1).
We can expect  enhanced proximity effect by the MARS.
In order to study this significantly novel charge transport,
we must construct a novel theory 
for DN/TS junctions 
beyond preexisting ones\cite{PRL2003,PRB2004,Nazarov2,Nazarov1}.
This is in fact very timely since triplet superconductors have
been discovered successively very recently \cite{Maeno}.
\par
\par
%
%
%
In the present paper, we  derive  a
conductance formula for  DN/TS junctions 
based on the  
Keldysh-Nambu (KN) Green's function formalism \cite{PRL2003,PRB2004}.
%
%
The  total zero voltage resistance $R$ in the DN/TS junctions
is significantly  reduced by the enhanced proximity effect in the 
presence of the MARS. 
At the same time,  local density of states (LDOS) 
in the DN region has zero energy peak (ZEP)
due to the penetration of the MARS into the DN region 
from the triplet superconductor (TS) side of the DN/TS interface. 
It is remarkable that when $R_{D}$ is sufficiently
larger than the Sharvin resistance $R_{0}$,
$R$ is  given by $R=R_{0}/C_{-}$,
which can become much smaller than the preexisting
lower limit value of $R$, $i.e.$,
$R_{0}/2 +R_{D}$.
In the above, $C_{-}$ is a constant completely
independent of both $R_{D}$ and $R_{B}$, where $R_{B}$ denotes the
interface resistance in the normal state.
When all quasiparticles injected at the interface feel the
MARS, $R$ is reduced to be
$R=R_{0}/2$ irrespective of the  magnitude of $R_{D}$ and $R_{B}$.
The line shape of the bias voltage $V$ dependent
conductance $\sigma(eV)$ has a giant ZBCP.
These novel features have never been expected
either in 
DN/CSS or DN/USS junctions.
\par
We consider a DN/TS junction
with  TS terminal
and normal reservoir (N)
connected by a quasi-one-dimensional diffusive
conductor (DN) having a resistance $R_{D}$.
The flat interface between the DN and the TS
has a resistance $R_{B}$ while the DN/N interface has zero
resistance.
The positions of the DN/N interface and the DN/TS
interface are denoted as $x=-L$ and $x=0$,
respectively as shown in Ref. \cite{PRB2004}. %
We restrict our attention to triplet  superconductors
with $S_{z}=0$ that preserves time reversal symmetry.
$S_{z}$ denotes the
$z$ component of the total spin of a Cooper pair.
It is by no means easy
to formulate a charge transport of DN/TS junctions 
since  the quasiparticle
Green's function has no angular dependence by the impurity scattering in 
the DN. 
However, as shown in our previous paper \cite{PRL2003,PRB2004},
if we concentrate on the matrix currents
\cite{Nazarov1,Nazarov2} via the TS to or from the DN,
we can make a boundary condition of the KN Green's function. 
We assume that the constriction area  between the DN and the TS is 
subdivided into an anisotropic zone in the DN,
two ballistic zones in the DN and the TS, and a scattering
zone, where both ballistic and diffusive regimes can be
covered \cite{PRB2004}.
The sizes of the ballistic zone in the DN
and the scattering zone in the current flow direction are much
shorter than the coherence length \cite{PRL2003,PRB2004,Nazarov2}.
%
%
%
The scattering zone is
modeled by an insulating delta function
barrier with the transparency
$T(\phi)=4\cos^{2}\phi/(4\cos^{2}\phi + Z^{2} )$,
where $Z$ is a dimensionless constant and $\phi$ is
measured from the interface normal to the junction \cite{PRB2004}.
%
The boundary condition for the KN Green's function in the DN
[$\check{ G}_{N}(x)$] at the DN/TS interface is
given by,
\begin{equation}
\frac{L}{R_{D}} \left.\left[\check{ G}_{N}(x) \frac{ \partial 
\check{G}_{N}(x) }{\partial x}%
\right]\right|_{x=0_{-}} =\frac{-h}{2e^{2} R_{B}} \langle \check{I} 
\rangle,  \label{Nazarov}
\end{equation}
using matrix current $\check{I}$ \cite{PRL2003,PRB2004}.
Average over the angle of injected particles at the interface
is defined by
\begin{equation}
\langle \check{I}(\phi) \rangle \equiv \int_{-\pi/2}^{\pi/2}\!\!\! d\phi\; 
\cos\phi
\, \check{I}(\phi)
\Big/ \int_{-\pi/2}^{\pi/2}\!\!\! d\phi\;  T(\phi)\cos\phi
\end{equation}
with $\check{I}(\phi)=\check{I}$.
The resistance of the interface $R_{B}$
is given by $R_{B}=<R_{0}>$.
%
As shown in the eq. (2) in our previous paper \cite{PRL2003},
matrix current $\check{I}$ is
a function of $T(\phi)$, $\check{G}_{1}=\check{G}_{N}(x=0_{-})$,
$\check{G}_{2+}$, and  $\check{G}_{2-}$,
where $\check{G}_{2+}$ ($\check{G}_{2-}$)
denotes the outgoing  (incoming)  Green's function
in the TS.
The Green's functions are fixed in the "TS" terminal and in the "N" terminal,
and the voltage $V$ is applied to the "N" terminal located at $x=-L$.
$\check{G}_{N}(x)$ is  determined from the
Usadel equation with eq. (1).
If we denote the retarded part of $\check{G}_{N}(x)$ and $\check{G}_{2\pm}$
as $\hat{R}_{N}(x)$ and $\hat{R}_{2\pm}$ \cite{PRB2004},
the following equations are satisfied,
$\hat R_{N}(-L) = \hat{\tau}_z$,
$\hat R_{2\pm}=(f_{\pm} \hat{\tau}_y + g_{\pm} \hat{\tau}_z)$
with $f_{\pm}=\Delta_{\pm}(\phi)/\sqrt{\Delta_{\pm}^{2}(\phi)-\epsilon^2}$
and  $g_{\pm}=\epsilon/\sqrt{\epsilon^2 - \Delta_{\pm}^{2}(\phi)}$,
using the Pauli matrices.
$\epsilon$ denotes the energy of the quasiparticles
measured from the Fermi energy.
$\Delta_{+}(\phi)$ [$\Delta_{-}(\phi)$] is the pair potential
felt by the outgoing (incoming) quasiparticles (see Fig. 1).
%
%
%
After some algebra,
we can show that $\hat{R}_{N}(x)$ is given by
$\sin\theta(x)    \cdot \hat{\tau}_x
+ \cos \theta(x) \cdot \hat{\tau}_z$.
%
%
The spatial
dependence of $\theta (x)$ in the DN
is determined by the following equation
\begin{equation}
D\frac{\partial ^{2}}{\partial x^{2}}\theta (x)+2i\epsilon \sin [\theta
(x)]=0,  \label{Usa1}
\end{equation}%
with diffusion constant $D$ in the DN.
%
Taking the
retarded part of Eq.~(\ref{Nazarov}), we obtain
\begin{equation}
\left. \frac{L}{R_{D}}\frac{\partial \theta (x)}{\partial 
x}\right|_{x=0_{-}}=\frac{%
\langle 2B_{R} \rangle}{R_{B}},
\label{b1}
\end{equation}
\[
B_{R} = \frac{ (\Gamma_{1} \cos\theta_{0} - \Gamma_{2} 
\sin\theta_{0})T(\phi) } { (2-T(\phi))\Gamma_{3}
+ T(\phi)[ \cos \theta_{0} \Gamma_{2} + \sin \theta_{0} \Gamma_{1} ] },
\]
with $\theta_{0}=\theta(x=0_{-})$,
$\Gamma_{2}=g_{+} +g_{-}$, $\Gamma_{3}=1 + f_{+}f_{-} + g_{+}g_{-}$
%
and $\Gamma_{1}=i(f_{+}g_{-} - g_{+}f_{-})$.
Here,  we focus on  $\epsilon=0$, where
the left hand side of Eq. (\ref{b1}) is reduced to be
$\theta_{0}/R_{D}$.
We define $F_{\pm}(\phi)$ as
$F_{\pm}(\phi)=\lim_{\epsilon \rightarrow 0} f_{\pm}
={\rm sign}(\Delta_{\pm}(\phi))$
with ${\rm sign}(\Delta_{+}(\phi))=1(-1)$ for $\Delta_{+}(\phi)>0(<0)$.
In the following, we define the terminology as follows :
when $\Delta_{+}(\phi)\Delta_{-}(\phi)>0$ is satisfied,
quasiparticles are in the {\it conventional channels} (CC),
while when $\Delta_{+}(\phi)\Delta_{-}(\phi)<0$ is satisfied,
quasiparticles are in the {\it unconventional channels} (UC).
In UC, quasiparticles feel the MARS
while in CC quasiparticles do not feel the
MARS.
Following the above definition,
$F_{+}(\pm \phi)=F_{-}(\pm \phi)$ is satisfied for CC,
while for UC,
$F_{+}(\pm \phi)=-F_{-}(\pm \phi)$ is satisfied.
 From eq. (4), we can show that
$B_{R}$ is zero for CC and $B_{R}=iF_{+}(\phi)$ for UC, respectively.
Then, $\theta_{0}$ becomes
\begin{equation}
\theta_{0}=iR_{D}C_{-}/R_{0}, \ C_{-}=\int_{UC} F_{+}(\phi) \cos \phi d\phi,
\end{equation}
where $\int_{UC}$ means the $\phi$ integral only from the UC within
$ -\pi/2<\phi <\pi/2$.
A remarkable feature is that $\theta_{0}$ becomes a purely imaginary
number as shown below.
 From eq. (3),
$\theta(x)$ at $\epsilon=0$ becomes
$\theta(x)=(x+L)\theta_{0}/L$.
Since the LDOS of the quasiparticles in the DN region renormalized  
by its value in normal state
is given by $\rho(\epsilon)={\rm Real}[\cos \theta(x)]$,
$\rho(0)$ is always larger than unity
except for $x=-L$.
This means that 
$\rho(0)$ is enhanced due to  the 
enhanced proximity effect by the MARS and 
has a ZEP as shown later.  
In the preexisting theories of DN/CSS and DN/USS junctions,
$\theta_{0}$ at $\epsilon=0$ is always a real number and
$\rho(x)$ never exceeds unity. \par
%
%
%

In order to understand the essential difference between the DN/TS junctions
and the DN/USS junctions intuitively, 
we consider four simplified cases,
(a)the DN/CSS junctions with  the CC,
(b)the DN/CSS junctions with  the UC,
(c)the DN/TS junctions with  the CC, and
(d)the DN/TS junctions with  the UC  for all $\phi$.
This situation is actually realized
by choosing  $d$-wave pair potential  with
$\Delta_{\pm}(\phi) = \Delta_0 \cos[2(\phi \mp \alpha)]$ (Figs. 1(a) and 1(b))
and $p$-wave pair potential with
$\Delta_{\pm}(\phi) = \pm \Delta_0 \cos(\phi \mp \alpha)$ (Figs. 1(c) and 1(d))
as a prototype of USS and TS, respectively.
%
The relations $F_{+}(\pm \phi)=F_{-}(\mp \phi)$
and $F_{+}(\pm \phi)=-F_{-}(\mp \phi)$, are satisfied
for the DN/USS and the DN/TS junctions,  respectively 
both for the UC and for the CC.
For the DN/USS junctions with the CC,
since $F_{\pm}(\phi)=F_{\pm}(-\phi)$ is satisfied,
the contribution to $\theta_{0}$ is
not cancelled by angular averaging [Fig. 1(a)].
For the DN/USS junctions with the UC and the DN/TS junctions with the CC,
since $F_{\pm}(\phi)=-F_{\pm}(-\phi)$ is satisfied,
the contribution to $\theta_{0}$ is
cancelled by the angular average [Figs. 1(b) and 1(c)].
However, for the DN/TS junctions with UC,  
since $F_{\pm}(\phi)=F_{\pm}(-\phi)$ is satisfied,
the  contribution  to $\theta_{0}$ is non-zero [Fig. 1(d)].
It is remarkable that for the DN/TS junctions
the CC do  not contribute
to the proximity effect while UC do. \par

\par
\begin{figure}[hob]
\begin{center}
\includegraphics[width=7.5 cm,clip]{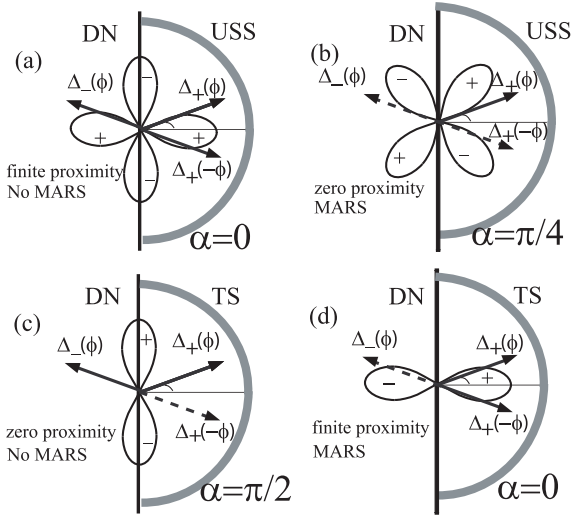}
\end{center}
\vskip -4mm
\caption{
Trajectories in the scattering process for  outgoing (incoming)
quasiparticles
for the DN/USS and the DN/TS junctions
and the corresponding  $\Delta_{\pm}(\phi)$ are schematically illustrated.
As a prototype, we choose $\Delta_{\pm}=\Delta_{0}\cos[2(\theta \mp 
\alpha)]$ for the
DN/USS junctions [(a) and (b)]
and $\Delta_{\pm}=\pm \Delta_{0}\cos[(\theta \mp \alpha)]$ for the DN/TS 
junctions
[(c) and (d)],  respectively,
where $\Delta_{0}$ is the maximum value of the pair potential.
The measure of the proximity effect $\theta_{0}$ is
determined by the integration for all injection angles, $i.e.$, the angular 
average over the hatched area.
}
\label{fig:01}
\end{figure}%
%
%
%
The total zero voltage resistance of the DN/TS junction
can be given as 
\begin{equation}
R=\frac{R_{D}}{L} \int^{0}_{-L}
\frac{dx}{\cos^{2} \theta(x)}
+ \frac{ R_{B}}{\langle I_{b0}\rangle},
\label{Rd}
\end{equation}
where $I_{b0}$ is obtained from the Keldysh part of
$\check{I}$
and is a complex function of
$g_{\pm}$, $f_{\pm}$, $\theta_{0}$, and $T(\phi)$.
%
%
At $\epsilon=0$,
$I_{b0}$ becomes
$1 + \exp(2\mid \theta_{0} \mid)$
for UC,
while for CC,
$I_{b0}$ becomes
$\frac{2T^{2}(\phi)}{[2-T(\phi)]^{2}}{\rm cos}^{2}\theta_{0}$.
After simple algebra,
%
$R$ is given by
\begin{equation}
R=R_{0}
\{
\frac{{\rm tanh}\theta_{0i} }{C_{-}}
+
\frac{2}{[1 + \exp(2\mid \theta_{0i} \mid)]C_{+}+{\rm cosh}^{2}\theta_{0i} D}
\}
\end{equation}
with $\theta_{0i}=-i\theta_{0}=R_{D}C_{-}/R_{0} $,
$C_{+}=\int_{UC}  \cos \phi d\phi$
and $D=\int_{CC}
\frac{2T^{2}(\phi)}{[2-T(\phi)]^{2}} d\phi$,
where $\int_{CC}$ means the $\phi$ integral only from the CC.
The resulting $R$ is  given by $R_{0}/C_{-}$
for sufficiently large $R_{D}/R_{0}$
and is independent both of
$R_{D}$ and $R_{B}$ except for the very special case with $C_{-}=0$.
%
The magnitude of $R$ can become much smaller than the 
preexisting lower limit value of $R$, $i.e.$, $R_{0}/2+R_{D}$.
This giant reduction of $R$ is due to the enhanced proximity effect by 
the MARS.
When all quasiparticles feel the MARS independent of $\phi$,
$C_{-}$ is given by 2 and $R$ becomes $R=R_{0}/2$
independent of $R_{D}$ and $R_{B}$.
This interesting situation is actually
realized for a $p_{x}$-wave case, 
where pair potentials are given by
$\Delta_{\pm}(\phi)=\pm\Delta_{0} \cos \phi$ [see case Fig. 1(d)].
Above enhanced proximity effect by the MARS is a completely 
novel proximity effect, which 
has never been  expected in any preexisting
theories \cite{Beenakker,PRL2003,PRB2004,Nazarov1,Nazarov2}.
\par
In order to understand this novel proximity effect in detail, 
we focus on the LDOS $\rho(\epsilon)$ in the 
DN region normalized by its value in the normal state.
We choose $p_{x}$-wave pairing as a prototype of the TS.
As a reference, we compare the results with those for 
the DN/USS junction with $d_{xy}$-wave pair
potential, where $\Delta_{\pm}(\phi)$ is given by
$\Delta_{\pm}(\phi)=\pm \Delta_{0}\sin(2\phi)$.
Although $\rho(\epsilon)$ at the  TS side of the 
DN/TS interface and  that at the USS side of the 
DN/USS interface both have ZEP by the formation of the MARS, 
$\rho(\epsilon)$ in the DN has a drastic
difference between the two cases. 
For the $p_{x}$-wave case, the LDOS in the DN region has a ZEP.
The height of the ZEP $\rho(0)$ is given by
$\cosh [2R_{D}(x+L)/(LR_{0})]$ and
the order of its  width is $E_{Th}$.
On the other hand, for the $d_{xy}$-wave case, $\rho(\epsilon)$
is always unity independent of
$R_{D}$ due to the absence of the proximity effect.
Contrary to the DN/TS junction, the MARS formed at the USS side of the 
DN/USS interface can not penetrate into the DN. 
As seen from $\rho(\epsilon)$ in DN/TS junctions, 
the significant reduction of $R$ 
originates from  the penetration of the MARS into the DN.
Although we have actually shown the existence of
ZEP of $\rho(\epsilon)$ in the DN for $p_{x}$-wave case as a prototype,
ZEP is universally expected for DN/TS junctions with the MARS at the interface 
independent of the detailed shape of the pair potentials of the TS.
On the other hand, the LDOS in the DN  region of DN/USS 
junctions do not have ZEP.
Using this clear and qualitative difference of LDOS in the DN region 
between the DN/TS and DN/USS junctions,
we can identify the triplet symmetry  of the pair potentials.
%
\par
\begin{figure}[htb]
\begin{center}
\includegraphics[width=5.10cm,clip]{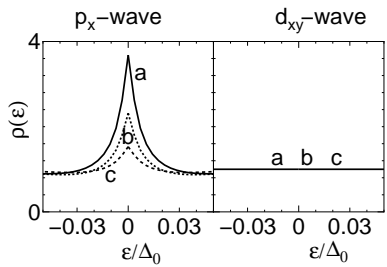}
\end{center}
\vskip -4mm
\caption{ $\rho(\epsilon)$ is plotted
  for the DN/TS junction with $p_{x}$-wave superconductor (left panel)
and the DN/USS junction with $d_{xy}$-wave superconductor (right panel)
for $Z=1.5$, $R_{D}/R_{B}=0.5$, and $E_{Th}=0.02\Delta_{0}$ with 
various $x$ in the DN.
a: $x=0_{-}$,
b: $x=-L/4$ and c:  $x=-L/2$.
}
\label{conductance}
\end{figure}
Finally, we focus on the line shape of the tunneling conductance
of DN/TS junctions for non zero voltage, which is defined by
$\sigma(eV)=R_{B}/R(eV)$.
We choose $p_{x}$-wave pairing as a prototype.
As a reference, we compare the results with those for the 
DN/USS junction with $d_{xy}$-wave pair potential.
Although both $p_{x}$-wave  and $d_{xy}$-wave cases
have similar line shapes of the voltage-dependent conductance
with the ZBCP as a function of $eV$  for ballistic junctions \cite{Yama},
$i.e.$ $R_{D}=0$ case,
we can classify these two for $R_{D} \neq 0$ as shown
in Fig. \ref{conductance}.
We choose the Thouless energy $E_{Th}$
as $E_{Th}=0.02\Delta_{0}$.
For the $p_{x}$-wave case,  since
all injected quasiparticles feel the MARS,
$\sigma(0)=2R_{B}/R_{0}$ is satisfied for any $R_{D}$.
For $R_{D} \neq 0$, $\sigma(eV)$ can be expressed by
the summation of the broad ZBCP and the narrow one, 
where $\sigma_{N}=R_{0}/R_{B}$
denotes the  angular averaged transparency of the junction.
The width of the former one is
proportional to $\sigma_{N}\Delta_{0}$ and the latter one is the
Thouless energy.
However, with the increase of $R_{D}/R_{B}$,
$\sigma(eV)$ for $\mid eV \mid >E_{Th}$ is suppressed,
and the  ratio of  $\sigma(0)$ to its   background value
is largely enhanced.
We can call this largely enhanced $\sigma(0)$ as
giant ZBCP (curve $b$ or $c$ in the left panel of Fig. 3).
By contrast,  for $d_{xy}$-wave case, since
$\sigma(0)=2R_{B}/(R_{0} + 2R_{D})$ is satisfied,
$\sigma(0)$ is reduced with the increase of
$R_{D}$.
The width of ZBCP is proportional to
$\Delta_{0}\sigma_{N}$ and is not changed
with the increase of $R_{D}/R_{B}$
due to the absence of the proximity effect.
%
%
\par

\begin{figure}[htb]
\begin{center}
\includegraphics[width=5.10cm,clip]{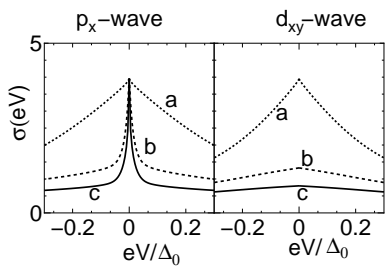}
\end{center}
\vskip -4mm
\caption{ $\sigma(eV)$ is plotted as a function
of $eV$ for the DN/TS junctions with $p_{x}$-wave superconductor (left panel)
and the DN/USS junctions with $d_{xy}$-wave superconductor (right panel)
for $Z=1.5$ and $E_{Th}=0.02\Delta_{0}$.
a: $R_{D}/R_{B}=0$,
b: $R_{D}/R_{B}=0.5$ and c:  $R_{D}/R_{B}=1$.
}
\label{conductance}
\end{figure}
%
%
In conclusion, we have presented a theory of charge transport in the DN/TS
junctions.
The total resistance $R$
%
can become much smaller than the
preexisting lower limit value of $R$ for DN/CSS and DN/USS
junctions, $i.e.$, $R_{0}/2+R_{D}$.
As the extreme case where all injected electrons
feel the MARS, $R$ is reduced to be $R_{0}/2$.
The significant reduction of $R$ is due to the 
enhanced proximity effect by  the MARS and the resulting LDOS in the DN region 
has  a ZEP.
At the same time, we can expect giant ZBCP.
The above enhanced proximity effect is 
a novel proximity effect, which has never been expected in
DN/CSS and DN/USS junctions.
We have shown in the present paper that these effects are
actually realizable for DN/TS junctions with
$p$-wave pair potential.
%
%
%
%
We believe that these features are easily verifiable in experiments since
a mesoscopic interference effect due to the
proximity effect  has recently
been observed in 
high $T_{C}$ cuprate junctions \cite{Hiromi}.
Similar experiments are technologically possible 
for junctions composed of  Sr$_2$RuO$_4$,  where triplet 
pairing is believed to be realized \cite{Maeno}. 
\par
%
The authors appreciate useful and fruitful discussions
with  Y. V. Nazarov, A. A. Golubov  J. Inoue, Y. Asano 
and K. Kuroki.
%
%
%


\begin{thebibliography}{99}
%
\bibitem{meso}
Y. Imry, 1996 {\it Introduction to mesoscopic Physics }
(Oxford University, London).
%


\bibitem{Beenakker}
C.W.J. Beenakker, Rev. Mod. Phys. {\bf 69} 731 (1997).


\bibitem{Hekking}
F. W. J. Hekking and Yu. V. Nazarov,
Phys. Rev. Lett. {\bf 71} 1625 (1993);
A.I. Larkin and Yu. V. Ovchinikov, Sov. Phys. JETP {\bf 41}
960 (1975);
A.F. Volkov, A.V. Zaitsev and T.M. Klapwijk,
Physica C 210 21 (1993).
%
%
\bibitem{TK95}
Y. Tanaka and S. Kashiwaya,
Phys. Rev. Lett. {\bf 74}, 3451 (1995).
%
\bibitem{Kashi00}
S. Kashiwaya and Y. Tanaka,
Rep. Prog. Phys. {\bf 63}, 1641 (2000);
The origin of the MARS and its relevance to ZBCP are discussed
in detail in this review.





\bibitem{Buch}
L.J. Buchholtz and G. Zwicknagl, Phys. Rev. B
{\bf 23} 5788 (1981); C. Bruder, Phys. Rev. B {\bf 41} 4017 (1990);
C.R. Hu,
Phys. Rev. Lett. {\bf 72}, 1526 (1994).
%


%
%
%
\bibitem{PRL2003}
Y. Tanaka, Yu. V. Nazarov and S. Kashiwaya
Phys. Rev. Lett. {\bf 90} 167003 (2003).
%
\bibitem{PRB2004}
Y. Tanaka, $et$ $al$, Phys. Rev. B  {\bf 69}  (2004), 
cond-mat/0311523.
The detailed derivation of a general form of
the matrix current are shown in this paper.
%



%
\bibitem{Nazarov2}
Yu. V. Nazarov, Superlatt. Microstruct. {\bf 25} 1221 (1999),
cond-mat/9811155.
%
\bibitem{Nazarov1}
Yu. V. Nazarov, Phys. Rev. Lett. {\bf 73} 1420 (1994).
%
\bibitem{Maeno}
A. P. Mackenzie and Y. Maeno, Rev. Mod. Phys. {\bf 75}
657 (2003);
Z.Q. Mao, $et$ $al$,
Phys. Rev. Lett. {\bf 87}, 037003 (2001);
Ch. W\"{a}lti,$et$ $al$,
Phys. Rev. Lett. {\bf 84}, 5616 (2000)
%
\bibitem{Yama}
M. Yamashiro $et$ $al.$, J. Phys. Soc. Jpn. {\bf 68} 3224 (1998);
C. Honerkamp and M. Sigrist J. Low. Temp. Phys. {\bf 111} 895 (1998).
%
%
\bibitem{Hiromi}
H. Kashiwaya,$et$ $al$, Phys. Rev. B {\bf 68} 054527 (2003).
\end{thebibliography}
\end{document}